\documentclass{nature}

\usepackage{amsmath}
\usepackage[pdftex]{graphicx}
\usepackage{dcolumn}  
\usepackage{bm}       
\usepackage{dsfont}
\usepackage{color}
\usepackage{braket}

\newcounter{defcounter}
\usepackage{amsthm}

\title{Unraveling Quantum Coherences Mediating Primary Charge Transfer Processes in Photosystem II Reaction Center } 

\author{Ajay Jha$^{1,2,3,4,*}$, Pan-Pan Zhang$^{1,*}$, Vandana Tiwari$^{2,5,*}$, Lipeng Chen$^{6}$, Michael Thorwart$^{7,8}$, R. J. Dwayne Miller$^{9}$, Hong-Guang Duan$^{1,2,7,8}$ }  

\begin{document} 

\maketitle 

\begin{affiliations}
\item Department of Physics, School of Physical Science and Technology, Ningbo University, Ningbo, 315211, P.R. China
\item Max Planck Institute for the Structure and Dynamics of Matter, Luruper Chaussee 149, 22761, Hamburg, Germany
\item Rosalind Franklin Institute, Harwell, Oxfordshire OX11 0QX, United Kingdom
\item Department of Pharmacology, University of Oxford, Oxford OX1 3QT, United Kingdom 
\item Department of Chemistry, University of Hamburg, Martin-Luther-King Platz 6, 20146 Hamburg, Germany
\item Zhejiang Laboratory, Hangzhou 311100, P.R. China 
\item I.\ Institut f\"ur Theoretische Physik,  Universit\"at Hamburg, Notkestr.\ 9, 22607 Hamburg, Germany 
\item The Hamburg Center for Ultrafast Imaging, Luruper Chaussee 149, 22761 Hamburg, Germany 
\item The Departments of Chemistry and Physics, University of Toronto, 80 St. George Street, Toronto Canada M5S 3H6 \\
$^*$These authors contributed equally to this work. \\ 
\centerline{\underline{\date{\bf \today}}} 
\end{affiliations} 

\begin{abstract} 

Photosystem II (PSII) reaction center is a unique protein-chromophore complex that is capable of efficiently separating electronic charges across the membrane after photoexcitation. In the PSII reaction center, the primary energy- and charge-transfer (CT) processes occur on comparable ultrafast timescales, which makes it extremely challenging to understand the fundamental mechanism responsible for the near-unity quantum efficiency of the transfer. Here, we elucidate the role of quantum coherences in the ultrafast energy and CT in the PSII reaction center by performing two-dimensional (2D) electronic spectroscopy at the cryogenic temperature of 20 K, which captures the distinct underlying quantum coherences. Specifically, we uncover the electronic and vibrational coherences along with their lifetimes during the primary ultrafast processes of energy and CT. We also examine the functional role of the observed quantum coherences. To gather further insight, we construct a structure-based excitonic model that provided evidence for coherent energy and CT at low temperature in the 2D electronic spectra. The principles, uncovered by this combination of experimental and theoretical analyses, could provide valuable guidelines for creating artificial photosystems with exploitation of system-bath coupling and control of coherences to optimize the photon conversion efficiency to specific functions. 

\end{abstract} 


The transfer of electrons in the PSII reaction center is one of the key mechanisms of converting sunlight energy into chemical energy within a cascade of photosynthetic reactions. The protein structure of the PSII reaction center has been well resolved by X-ray crystallography \cite{RC_crystal, RC_crystal2, RC_crystal3}. It contains eight cofactors symmetrically located along the protein central axis. Its protein structure  is shown in Fig.\ \ref{fig:Fig1}(a), which includes two primary chlorophylls $\rm P_{D_{1}}$ and $\rm P_{D_2}$, two accessory chlorophylls $\rm Chl_{D_1}$ and $\rm Chl_{D_2}$, two pheophytins $\rm Pheo_{D_1}$ and $\rm Pheo_{D_2}$, and two peripheral chlorophylls $\rm Chlz_{D_1}$ and $\rm Chlz_{D_2}$. Recent studies have shown that only the $\rm D_{1}$ branch is active in the charge separations in the reaction center \cite{D1_branch}. In 2005, Novoderezhkin {\em et al.} developed an exciton model and refined parameters by a simultaneous fit to the absorption, emission, linear dichroism (LD) and circular dichroism (CD) spectra \cite{Novoderezhkin_I}. The model and the features of the CT states have been further improved by fitting the model to the additional steady-state fluorescence and Stark spectra \cite{Novoderezhkin_II}. Then, photo-echo and transient absorption spectroscopies were performed to measure the primary processes of CT in the PSII reaction center \cite{Valentyn}. It has been suggested that the primary CT ($\rm Chl_{D_{1}} Pheo_{D_{1}})^{*} \rightarrow Chl^{+}_{D_{1}} Pheo^{-}_{D_{1}}$ occurs on the timescale of 1.5 ps, while the secondary charge transfer process ($\rm Chl^{+}_{D_{1}} Pheo^{-}_{D_{1}} \rightarrow P^{+}_{D_1} Pheo^{-}_{D_1}$) occurs within 25 ps. Due to a relatively slow reaction process of the secondary charge transfer, the PSII reaction center was further investigated by transient absorption spectroscopy with an extension of the detection window up to 3 ns \cite{ChemPhysChem, Biochemistry}. The measured data were examined by global and target analyses and show four CT states participating in two CT processes within the timescales ranging from hundreds of femtoseconds to nanoseconds. 

The recently developed 2D electronic spectroscopy is one of the powerful tools to study the ultrafast energy and CT in photosynthetic protein complexes \cite{Engel2007, Scholes2010, Duan_Sci_Adv2020, Duan2017, Scholes2017, Zigmantas2018, Duan2022}. In the 2D spectrum, the unique separation of excitation and detection windows allows us to resolve the homogeneous signals from the ensemble-averaged nonlinear optical responses. Moreover, this separation enables us to directly monitor the energy and the CT and the associated coherent dynamics as a function of absorption and emission wavelengths \cite{Nature_rev_chem2019, Romero2017}. These features have been assumed to be particularly useful in examining rather complex systems with congested manifolds of electronic states coupled to the vibrational modes \cite{David2013, Plenio2013, Miller2014, Duan2015, JPCL2012, Dassia2014, Hauer2013}. However, the absorption spectrum of the PSII reaction center manifests that the optical transitions of the 8 cofactors mainly fall within a range of 40 nm in the visible, i.e., from 640 to 680 nm.  It is this particular feature which makes it extremely challenging to effectively resolve all components by any spectroscopic approach and to interpret the congested line shapes. The first 2D spectroscopic study of the PSII reaction center has been performed by Myers {\em et al.} at 77 K \cite{Jennifer2010}. They uncovered several timescales of the relevant energy and CT processes on the basis of fitting of each individual trace in the 2D spectra along the waiting time. A corresponding model based on observed parameters has been constructed in Ref.\ \cite{Jennifer2012} and further refined by a tight-binding model with 5 particular CT states \cite{Jennifer2013}. Due to the lack of a direct transition dipole moment for CT states, it has been shown that the spectroscopic signature of CT dynamics is challenging to observe even within 2D electronic spectra. Moreover, Fuller {\em et al.} and Romero {\em et al.} independently examined the role of electronic and vibrational coherences by 2D electronic spectroscopy \cite{Romero2014, Jennifer2014} at 77 K. Both studies suggested that the CT dynamics in the PSII reaction center is supported by vibronic coherence, especially when the frequencies of the oscillations are resonant with the excitonic energy gap. However, due to the spectroscopic congestion, the featureless 2D spectra failed to resolve the excitonic site energies and the cross peaks at 77 K, which resulted in a disagreement for the excitonic energies reported in both works with respect to modeling of the PSII reaction center. Although the CT dynamics in the PSII reaction center have been further studied by multicolor 2D electronic spectroscopy and the 2D electronic-vibrational spectroscopy \cite{Fleming2022, Jennifer2021}, obviously, it is still elusive to obtain even conclusive excitonic energies for the PSII reaction center, let alone the assignment of a spectroscopic basis of the electronic quantum coherence for a particular site of a cofactor. 

Here, we study the coherent dynamics of the energy and CT processes in the PSII reaction center by 2D electronic spectroscopy. We aim to capture solid evidence of the electronic and vibrational quantum coherences by performing the spectroscopic measurement at a low temperature, namely 20 K. At this temperature, we are able to clearly resolve the lifetime of the electronic coherence and the cross peaks in the 2D electronic spectra. The data analysis and the associated theoretical calculations enable us to identify the coherent components in the primary CT of  $\rm Chl_{D_{1}}$, $\rm Pheo_{D_1}$, $\rm P_{D2}$ and $\rm P_{D1}$. We show that the lifetime of the electronic coherence is 180 fs, which will persist at a rapidly decaying amplitude out to 600 fs, corresponding to the primary CT of $\rm (Chl_{D1}Pheo_{D1})^{*} \rightarrow Chl^{+}_{D1}Pheo^{-}_{D1}$. Moreover, we unravel that the coherent CT in radical pairs ($\rm P_{D1}$, $\rm P_{D2}$) occurs on the timescale of 180 fs. We further construct an excitonic model and refine the parameters by simultaneously fitting them to the experimental spectra. The system-bath interaction and the reorganization energy of the PSII reaction center is further confirmed by comparing to the anti-diagonal profile of the 2D spectrum at zero waiting time. By this information, our model is capable of capturing the correct lifetime of the electronic coherence in the primary processes of the CT. We further uncover that the strong excitonic coupling between $\rm P_{D1}$ and $\rm P_{D2}$ results in a robust coherent dynamics even at room temperature. In contrast, the electronic coherence between $\rm Chl_{D1}$ and $\rm Pheo_{D1}$ only survives at 20 K. With the combination of experimental and theoretical effects, we clarify the role of electronic coherence in the ultrafast processes of the CT in the PSII reaction center. Due to the rather large magnitude of the reorganization energies of the CT states, we do not observe any functional role of the quantum coherence in the processes of the primary CT even at 20 K. This low-temperature measurement provides a solid basis to examine the quantum effects in the energy and CT dynamics in natural photosynthetic protein complexes. The principles uncovered in this work could help in the rational design of artificial photovoltaic devices by providing the optimal system-bath coupling parameters for directing photochemical processes and any gains possible through conserved quantum coherences. 

\section*{Results} 

\begin{figure}[h!]
\begin{center}
\includegraphics[width=18.0cm]{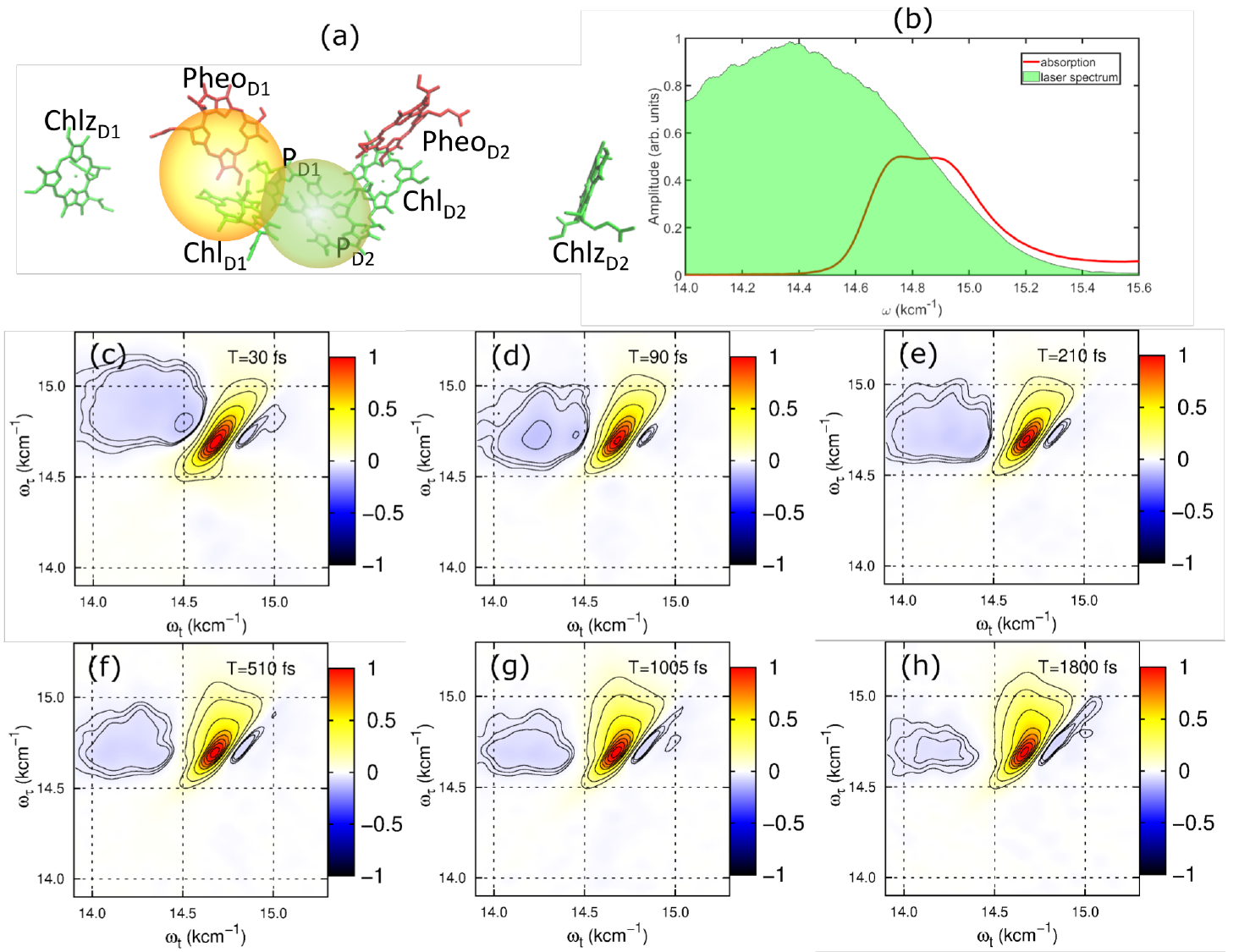}
\caption{\label{fig:Fig1} (a) Structural arrangements of the pigments in the PSII reaction center protein complex. (b) Absorption spectrum of the PSII reaction center at 80 K and the overlap of the laser spectrum used in this measurement. The real part (total) of the 2D electronic spectra of the RC complex measured at 20 K for selected waiting times  30, 90, 210, 510, 1005 and 1800 fs are shown from (c) to (h), respectively. The red (positive) peaks in the 2D spectra denote the GSB signal and SE, the blue (negative) magnitude indicates the contribution of ESA. } 
\end{center}
\end{figure}

The solution of the PSII reaction center is prepared in a home-built sample cell and mounted in the Oxford Instrument cryostat. More details of the sample preparation are given in the section of Materials and Methods. Fig.\ \ref{fig:Fig1}(a) presents the structural arrangement of the 8 chlorophylls in the protein matrix (the protein data have been obtained from 2AXT.pdb). The measured absorption spectrum of the PSII reaction center at 80 K is shown in Fig.\ \ref{fig:Fig1}(b), together with the laser spectrum in this measurement. We selectively shifted our laser spectrum to the red side of the absorption spectrum of the reaction center to avoid the excitation of high-frequency vibrations during the measurements. 

\subsection{Two-dimensional electronic spectroscopy} 

We measured the 2D electronic spectra of the PSII reaction center at 20 K. The details of 2D spectrometer are described in the Materials and Methods section. The real part of the 2D electronic spectra are shown in Fig.\ \ref{fig:Fig1}(c) to (h) for selected waiting times at 30, 90, 210, 510, 1005 and 1800 fs. In Fig.\ \ref{fig:Fig1}(c), the positive and negative amplitude of the 2D spectra are plotted as red and blue peaks. They indicate the ground-state-bleach (GSB), stimulated emission (SE) and the excited state absorption (ESA), respectively. We first show the measured 2D electronic spectrum for the waiting time T = 30 fs. We observe that the optical transitions of the PSII reaction center are mainly located in a range from 14500 to 15000 cm$^{-1}$. Moreover, we observe a dramatic stretch of the main peak along the diagonal, which is indicative of strong inhomogeneous broadening of the cofactors in the reaction center. In addition, one cross peak with negative amplitude is present in the  upper-left region of the  2D spectrum. At T = 90 fs, the 2D spectrum does not change significantly, except that the elongation of the main peak is slightly reduced. This elongation is further reduced for T = 210 and 510 fs. Meanwhile, the anti-diagonal bandwidth of the main peak increases with the waiting time and even so for the waiting times T = 1005 and 1800 fs. More importantly, we observe cross peaks at ($\omega_{\tau}$, $\omega_{t}$) = (15000, 14700) cm$^{-1}$. They manifest the down-hill energy and CT. Moreover, the cross peak of the ESA (blue peak) reduces its magnitude for increasing waiting times. It has been demonstrated that the time constant of the  electronic dephasing of the coherence between the ground and excited states can be directly measured by the anti-diagonal bandwidth in the 2D spectrum \cite{Duan2017}. By this, we retrieve the anti-diagonal profile of the main peak from the rephasing part of  the 2D spectrum at ($\omega_{\tau}$, $\omega_{t}$) = (14700, 14700) cm$^{-1}$. Based on the fit of a Lorentizan lineshape function, we determine the time constant of the electronic dephasing to be 165 fs at 20 K. Further details are described in the Supporting Information (SI). 

\subsection{Primary charge transfer and electronic quantum coherence } 

\begin{figure}[h!]
\begin{center}
\includegraphics[width=15.0cm]{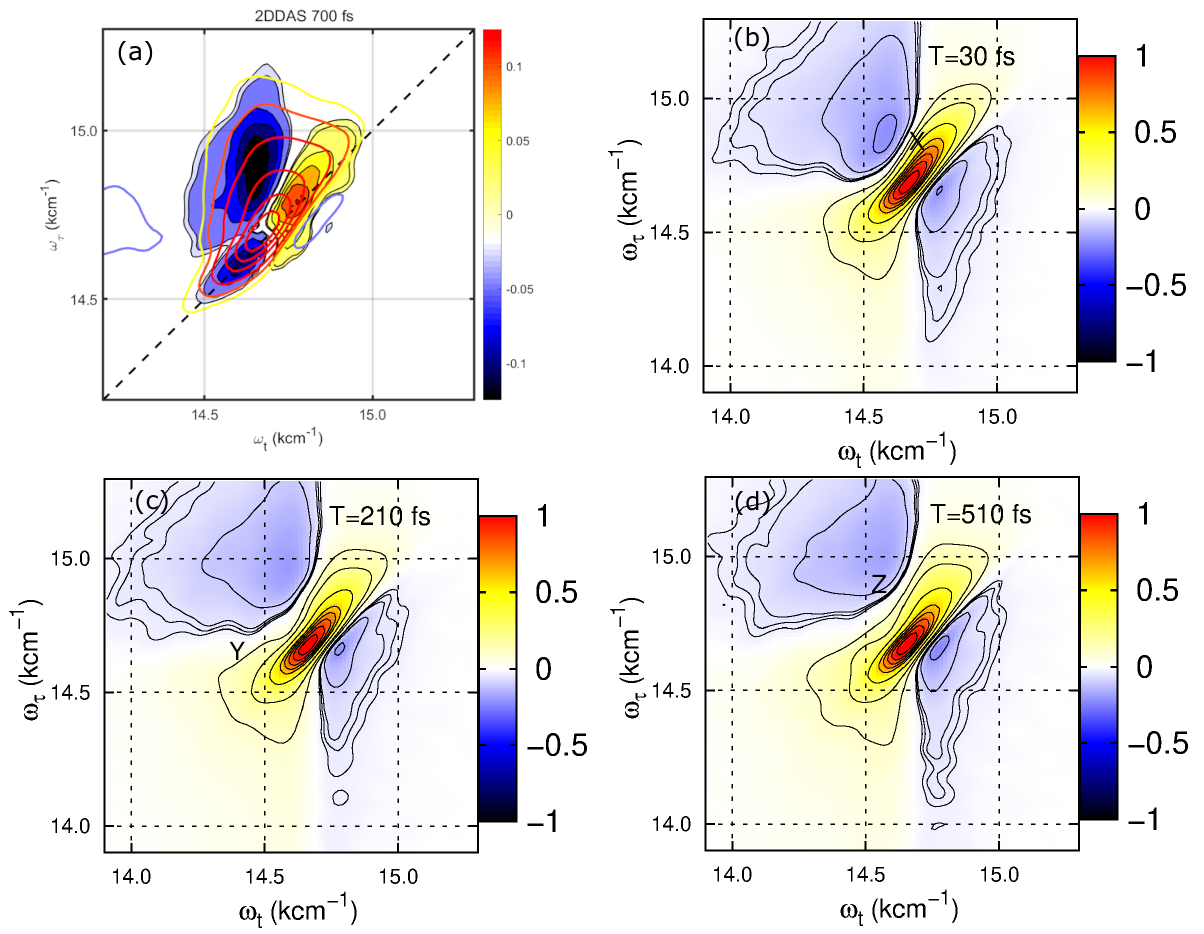}
\caption{\label{fig:Fig2} (a) 2DDAS with the decay time constant of 700 fs. This component reveals the ultrafast energy- and charge-transfer dynamics associated with the electronic quantum coherence. (b) Real part of the rephasing component of the 2D electronic spectrum at T = 30 fs. The selected cross peak is marked by ``X" with ($\omega_{\tau}$, $\omega_{t}$) = (14812, 14690) cm$^{-1}$. The rephasing part of the 2D spectra of 210 and 510 fs are presented in (c) and (d). The marked cross peak ``Y" shows the coordinate ($\omega_{\tau}$, $\omega_{t}$) = (14690, 14350) cm$^{-1}$. In addition, the marked cross peak ``Z" is located at  ($\omega_{\tau}$, $\omega_{t}$) = (14874, 14526) cm$^{-1}$. } 
\end{center}
\end{figure}

To study the coherent dynamics in the 2D spectra, we next construct a three-dimensional data set by resorting the time series of the 2D electronic spectra with evolving waiting time. We subsequently perform global fitting \cite{global fitting} to retrieve the decay-associated spectrum (DAS) with the separated time components. This is given by $\rm S(\omega_{\tau}, T, \omega_{t}) = \sum_{i}A_{i}(\omega_{\tau},\omega_{t}) \exp(-T/\tau_{i})$, where $\rm A_{i}(\omega_{\tau}, \omega_{t})$ is the DAS with the decay times $\rm \tau_{i}$. By this, we obtain six 2D DAS with the lifetime components of 27 fs, 700 fs, 4.9 ps, 27 ps, 104 ps and infinity, respectively. The obtained 2D DAS with the lifetime of 700 fs is shown in Fig.\ \ref{fig:Fig2}(a). It is superimposed with the  contours of the 2D electronic spectrum (T = 510 fs). The remaining components of the 2D DAS are shown in the SI. 

The fastest component of 27 fs manifests the peak broadening and the pulse-overlap effect in the 2D spectra at initial waiting time. The second component of 700 fs shows an interesting feature of diagonal and off-diagonal peaks, as shown in Fig.\ \ref{fig:Fig2}(a). In the DAS, we observe the diagonal peak with a positive magnitude from 14700 to 15000 cm$^{-1}$, which reveals the population decay at this wavelength. On the other hand, we observe a clear diagonal peak (negative) centered at 14600 cm$^{-1}$ in the DAS.  It  indicates the population increase in the 2D electronic spectra evolving with the waiting time. Moreover, the cross peak at ($\omega_{\tau}$, $\omega_{t}$) = (14800, 14600) cm$^{-1}$ shows a negative amplitude. This signature has been considered as  solid evidence of the down-hill population transfer from the positive diagonal to the negative diagonal peaks \cite{DuanJPCB2015, DuanSciRep_2017}. Interestingly, the associated lifetime of 700 fs of this DAS component is comparable to the timescale of coherent CT in the PSII reaction center. In addition, we also retrieve the 2D DAS with timescales of the 4.9 ps, 104 ps and infinity components. We expect that these components strongly relate to the secondary CT in the PSII reaction center. The detailed discussion of the 2D DAS and the data analysis are shown in the SI. Here, we mainly focus on the CT in the decay window of 700 fs, which show the timescale comparable to the coherent dynamics at 20 K. 

For this, we study the coherent dynamics in the rephasing part of the 2D electronic spectra. We show the real part of the rephasing 2D electronic spectra at T = 30, 210 and 510 fs in Fig.\ \ref{fig:Fig2}(b), (c) and (d), respectively. To examine the primary CT, we construct an excitonic model and refine our parameters by a careful fit to the experimental absorption and 2D spectra. The detailed description of the theoretical modeling and the parameters are shown in the next section. In our previous work, we have demonstrated the signature of the primary CT in the 2D DAS \cite{DuanSciRep_2017}. We unraveled two primary CT pathways  
\begin{eqnarray}
\label{eq:PCT}
&(Chl_{D1}Pheo_{D1})^{*} \rightarrow Chl^{+}_{D1} Pheo^{-}_{D1}, \\ 
&(P_{D1}P_{D2})^{*} \rightarrow P^{+}_{D2}P^{-}_{D1}. 
\end{eqnarray}
To examine the first primary CT, we resolve the cross peak at ($\omega_{\tau}$, $\omega_{t}$) = (14812, 14690) cm$^{-1}$, which has been marked as ``X" in Fig.\ \ref{fig:Fig2}(b). Based on our model, we identify that the excitonic states at 14812 and 14690 cm$^{-1}$ are associated to the pigments $\rm Chl_{D1}$ and $\rm Pheo_{D1}$. The details of the refined system Hamiltonian and the basis transformation from site to excitons are described in the SI. Thus, we expect that the associated coherent dynamics of both pigments are recorded in the frequency coordinates ($\omega_{\tau}$, $\omega_{t}$) = (14812, 14690) cm$^{-1}$ in the 2D electronic spectra. Moreover, we resolve the cross peak (marked as ``Y" in Fig.\ \ref{fig:Fig2}(c)) at ($\omega_{\tau}$, $\omega_{t}$) = (14690, 14350) cm$^{-1}$. It corresponds to the excitonic state ($\rm Chl_{D1}Pheo_{D1}$)$^{*}$ and the CT state ($\rm Chl^{+}_{D1} Pheo^{-}_{D1}$). 

\begin{figure}[h!]
\begin{center}
\includegraphics[width=10.0cm]{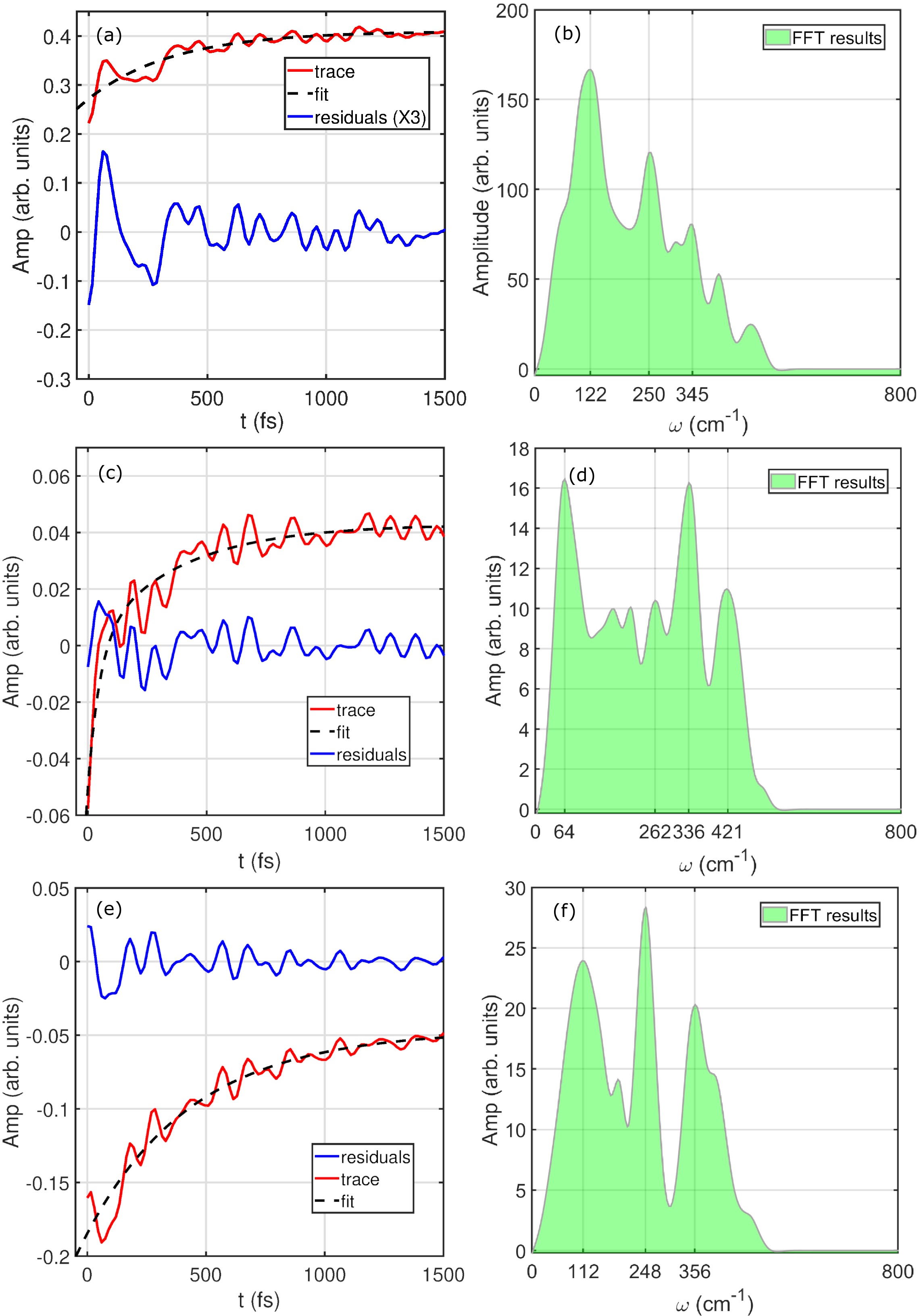}
\caption{\label{fig:Fig3} (a) Trace of the cross peak ``X" (red solid line). The global fitting curve (black dashed line) and residuals (blue solid line) are plotted in (a) as well. The residuals are magnified three times to reveal the oscillatory dynamics. The Fourier transform of the residuals are shown in (b) accordingly, with the resolved oscillation frequencies  122, 250 and 345 cm$^{-1}$, respectively. The traces of the selected cross peaks of ``Y" and ``Z" are shown in (c) and (e). The subsequent fitting results and residuals are plotted as black dashed lines and blue solid lines, respectively. The results of the Fourier transform are shown in (d) and (f) with the corresponding oscillation frequencies. } 
\end{center}
\end{figure}

To study the other pathway of the CT process, we identify the cross peak at ($\omega_{\tau}$, $\omega_{t}$) = (14874, 14526) cm$^{-1}$, which is marked as ``Z" in Fig.\ \ref{fig:Fig2}(d). From the basis transformation, we find that this cross peak is associated to the excitonic states ($\rm P_{D1}P_{D2}$)$^{*}$ and ($\rm P^{+}_{D2}P^{-}_{D1}$). To examine the coherent dynamics of these identified peaks, we extract the time traces of the cross peaks and show them in Fig.\ \ref{fig:Fig3}. In Fig.\ \ref{fig:Fig3}(a),  the raw data of the  trace is depicted as a red solid line. To reduce the noise level, we average a square of 25 pixels (5$\times$5) and plot the averaged values. We also perform the Tukey window Fourier transform to remove the high-frequency jitters during the measuring process. The detailed description of the data treatment by the Tukey window Fourier transform is given in the SI. We further subtract the kinetics by a fit with exponential functions (the global fitting approach is reviewed in the SI), which is plotted as black dashed line in Fig.\ \ref{fig:Fig3}(a). We then depict the obtained residuals as a blue solid line with a three-fold magnification to observe the clear oscillatory dynamics. The Fourier transforms of the residuals  are shown in Fig.\ \ref{fig:Fig3}(b). They show three well resolved peaks at the frequencies of 122, 250 and 345 cm$^{-1}$, respectively. 

We then extract the time-evolved trace of the marked peak ``Y" and plot it as red solid line in Fig.\ \ref{fig:Fig3}(c). The kinetics are fitted by exponential functions and the residuals are obtained as blue solid line. The results of the Fourier transform of the residuals are plotted in Fig.\ \ref{fig:Fig3}(d). They yield four  frequencies of 64, 262, 336 and 421 cm$^{-1}$, respectively. Moreover, we repeat the same procedure and plot the raw data of the trace and the residuals in Fig.\ \ref{fig:Fig3}(e). The subsequent results of the Fourier transforms   are shown in Fig.\ \ref{fig:Fig3}(f). They show the modes with the frequencies of 112, 248 and 356 cm$^{-1}$. These resolved modes coincide with the those of the experimental observations in Refs.\ \cite{Jennifer2014, Romero2014} and they also agree with previous low-temperature fluorescence line narrowing measurements \cite{PNAS1995}. 

\begin{figure}[h!]
\begin{center}
\includegraphics[width=9.0cm]{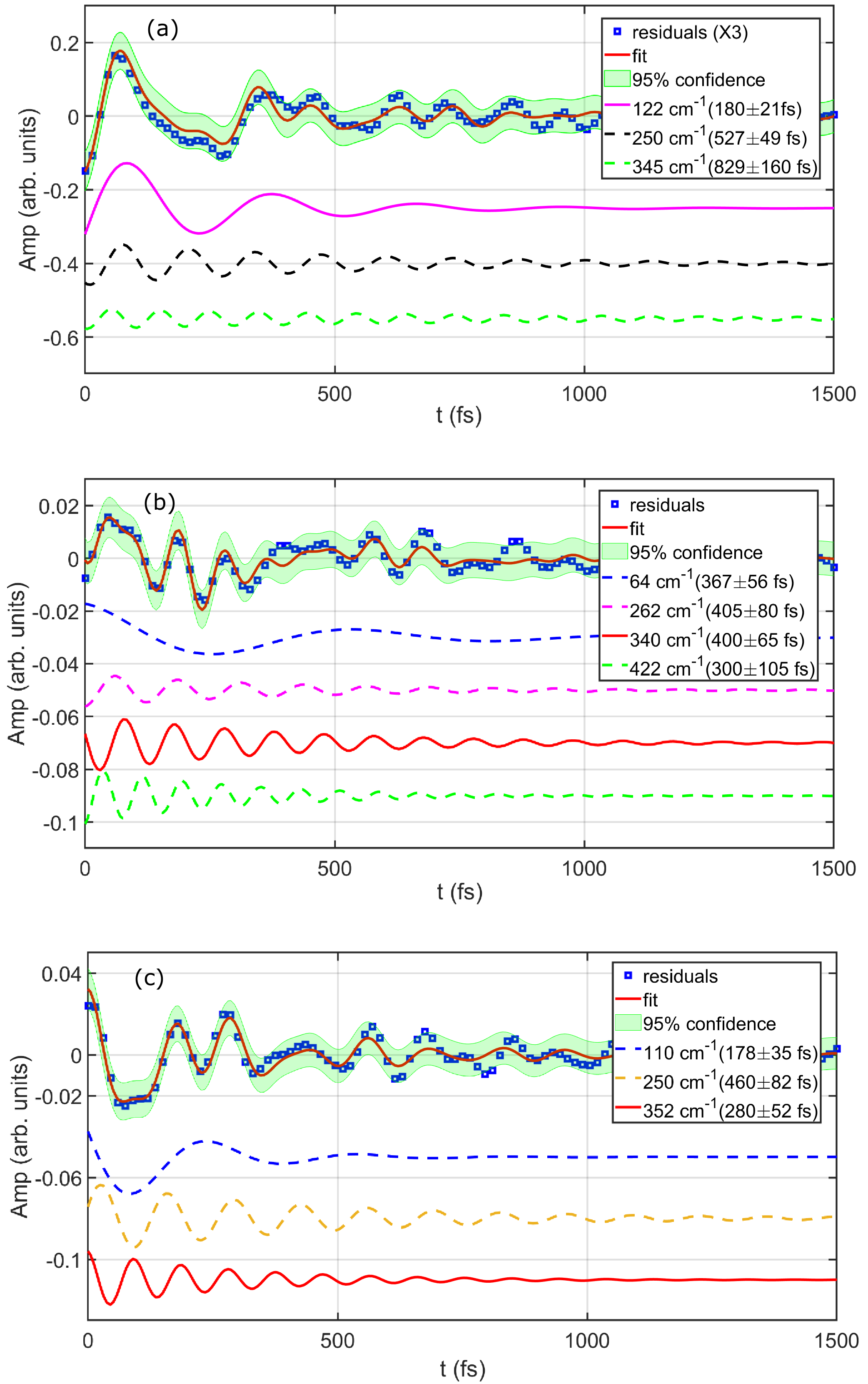}
\caption{\label{fig:Fig4} (a) Time trace of the cross peak ``X" (blue square dots). The fitting results are shown as red solid line with 95\% of confidence (green area). The resolved oscillatory dynamics are shown at the frequencies 122, 250 and 345 cm$^{-1}$, together with their lifetimes of 180, 527 and 829 fs, respectively. (b) Trace of the cross peak ``Y" (blue square dots). The fitting curve is plotted as red solid line with 95\% of confidence. The resolved oscillations occur with frequencies 64, 262, 340 and 422 cm$^{-1}$. Their lifetimes are 367, 405, 400 and 300 fs, respectively. (c) Trace of the cross peak ``Z" (blue square dots) and the fitting curve are shown as red solid line. The identified oscillations have the frequencies 110, 250 and 352 cm$^{-1}$, their lifetimes are extracted to be  178, 460 and 280 fs, respectively. } 
\end{center}
\end{figure}

We then proceed to disentangle the coherent dynamics and the associated lifetimes of each component. For this, we perform the data analysis of the residuals by fitting exponentially decaying harmonic functions in order to extract the oscillation frequencies and the lifetime of the coherences. The detailed fitting functions and procedures are shown in the SI. We first show the residuals of the cross peak ``X" in Fig.\ \ref{fig:Fig4}(a) as blue square dots. We start our fitting procedure with the resolved oscillatory frequencies: 122, 250 and 345 cm$^{-1}$. We then assign the initial values of the lifetime to be 300 fs with the guessed range from 0 to infinity. All the fitting procedures were performed using the Curve fitting Toolbox (Matlab2021(b)). We finalize the fitting results with the R-square $>$0.97. They are shown as red solid line in Fig.\ \ref{fig:Fig4}(a). The green shadow indicates the boundary  of the  confidence interval of 95\%. By this, we are able to separate the electronic coherence from vibrational coherence. The coordinates of cross peak ``X"  ($\omega_{\tau}$, $\omega_{t}$) = (14812, 14690) cm$^{-1}$ reveal the energy gap of 122 cm$^{-1}$ in the 2D electronic spectra. The matching of the resolved frequency of this mode and the energy gap manifests that the oscillatory dynamics of 122 cm$^{-1}$ corresponds to the electronic quantum coherence. Moreover,  theoretical calculations and the representation transformation from the site to the exciton basis reveal that  the cross peak ``X" mainly originates from the transition between $\rm Chl_{D1}$ and $\rm Pheo_{D1}$. Thus, this oscillatory component of 122 cm$^{-1}$ (magenta solid line in Fig.\ \ref{fig:Fig4}(a)) reveals  the electronic coherence between $\rm Chl_{D1}$ and $\rm Pheo_{D1}$. It shows the lifetime of 180 fs, which perfectly agrees with the lifetime retrieved from the theoretical modeling and the calculations discussed in the next section. In addition, we also resolve the vibrational coherence of frequencies 250 and 345 cm$^{-1}$ (black and green dashed lines in Fig.\ \ref{fig:Fig4}(a)).  They have the lifetimes of 527 and 829 fs, respectively. 

We repeat the same procedure to treat the data and analyze the coherent dynamics of the cross peaks ``Y" and ``Z". The residuals of peak ``Y" are plotted as blue squares in Fig.\ \ref{fig:Fig4}(b). The fitting procedure is performed and the fitting quality is illustrated by the red solid line with 95\% of confidence interval (green shadow). The resolved oscillations are plotted as dashed and solid lines,  respectively. We notice that the energy gap of the cross peak ``Y" (($\omega_{\tau}$, $\omega_{t}$) = (14690, 14350) cm$^{-1}$) is 340 cm$^{-1}$ and it coincides with the frequency mode of 340 cm$^{-1}$ (red solid line) in Fig.\ \ref{fig:Fig4}(b). The representation transformation between sites and excitons reveals that the cross peak ``Y" mainly originates from the exciton state ($\rm Chl_{D1}Pheo_{D1}$)$^{*}$ and the CT state ($\rm Chl^{+}_{D1}Pheo^{-}_{D1}$). However, our theoretical calculations show that the electronic coherence of the CT states is absent due to the strong dissipation induced by the permanent dipole moment of the CT states. The detailed calculations are presented in the next section. In addition, the mode of 340 cm$^{-1}$ also agrees with the resolved frequency of the vibrational mode found by other groups \cite{Jennifer2014, Romero2014, PNAS1995}. Thus, we infer that the resolved 340 cm$^{-1}$ is of vibrational origin and the data analysis yields a lifetime of 400 fs. Moreover, the other frequencies are resolved at 64, 262 and 422 cm$^{-1}$ in Fig.\ \ref{fig:Fig4}(b), respectively. 

The residuals of the cross peak ``Z" are shown as blue squares in Fig.\ \ref{fig:Fig4}(c). The fitting procedure is performed and  the quality of the fit is indicated by the red solid line with 95\% of confidence interval, together with the green shadow. We are able to discern  the frequencies of the modes at 110, 250 and 352 cm$^{-1}$, respectively. The energy gap of the cross peak ``Z" (($\omega_{\tau}$, $\omega_{t}$) = (14874, 14526) cm$^{-1}$) is 348 cm$^{-1}$, which is close to the resolved frequency of 352 cm$^{-1}$. On the basis of our calculations, we conclude that the cross peak ``Z" mainly originates from the excitonic state ($\rm P_{D1}P_{D2}$)$^{*}$ and $\rm P^{+}_{D2}P^{-}_{D1}$. Moreover, our calculations show that the electronic coherence of the CT state $\rm P^{+}_{D2}P^{-}_{D1}$ is quite short-lived, which is due to the strong system-bath interaction of the CT state. Based on this, we assign the resolved mode of 352 cm$^{-1}$ (red solid line in Fig.\ \ref{fig:Fig4}(c)) to the vibrational or, more precisely, to the  vibronic quantum coherence (see the 2D vibrational spectra of $\rm \omega_{T}=351$ cm$^{-1}$ in the SI). This mode also agrees with the modes identified in Ref.\ \cite{PNAS1995}. Additionally, on the basis of the fitting procedure, we also identify the other vibrational coherence with frequencies of 110 and 250 cm$^{-1}$ (blue and yellow dashed lines). They have the lifetimes of 178 and 460 fs, respectively. 

We have studied the coherent dynamics of particular frequency coordinates in the 2D electronic spectra. We next examine the coherent dynamics in the whole frequency map. For this, we consider the three-dimensional residuals within the global fitting procedure and perform the Fourier transform along the waiting time T to extract the oscillatory dynamics along $\rm \omega_{T}$. By this, we obtain the 2D vibrational maps $\rm S(\omega_{\tau}, \omega_{T}, \omega_{t})$ with the resolved frequencies $\rm \omega_{T}$. They are shown in the SI. We identify 5 components with the frequencies of 98, 117, 253, 351 and 741 cm$^{-1}$. All agree with the previous measurements of other groups \cite{Jennifer2014, Romero2014, PNAS1995}. To clarify the origin of the  vibrations, the 2D vibrational maps are plotted together with the contours of the 2D electronic spectrum. The cross peaks presented in the 2D vibrational maps show clear evidence of the vibrational progression \cite{Dassia2014}. The latter is marked as blue dashed lines with the frequency gap of $\rm \omega_{T}$. More details of 2D vibrational spectra are presented in the SI. 

\subsection{Theoretical calculations} 

We construct a tight-binding model to study the coherent dynamics of PSII reaction center. The electronic transitions in the pigments are approximated by optical transitions between the LUMO and HOMO. In addition, the electronic couplings between pigments are calculated within the dipole approximation. To take quantum dissipation into account, we assume that each pigment is linearly coupled to its own thermal reservoir. We use a standard Ohmic spectral density to examine the role of the electronic, vibrational and vibronic quantum coherences and their interplay. The 2D electronic spectra of the PSII reaction center are calculated using response function theory. The time propagation is calculated by the modified Redfield quantum master equation \cite{Duan_PRE_2015, JCP1998, Fleming2002}. More details are given in the Materials and Methods section and the SI. We use the values of the site energies and electronic couplings of the pigments initially from previous works \cite{Jennifer2013, ChemPhysChem} and optimized then the site energies by simultaneously fitting to the measured absorption spectra at different temperatures. After that, we calculate the 2D electronic spectra and refine the system-bath coupling strengths by comparing the calculated electronic dephasing lifetimes to the measured ones at different temperatures. By this, we obtain the optimized set of parameters for the system-bath model.

\begin{figure}[h!]
\begin{center}
\includegraphics[width=15.0cm]{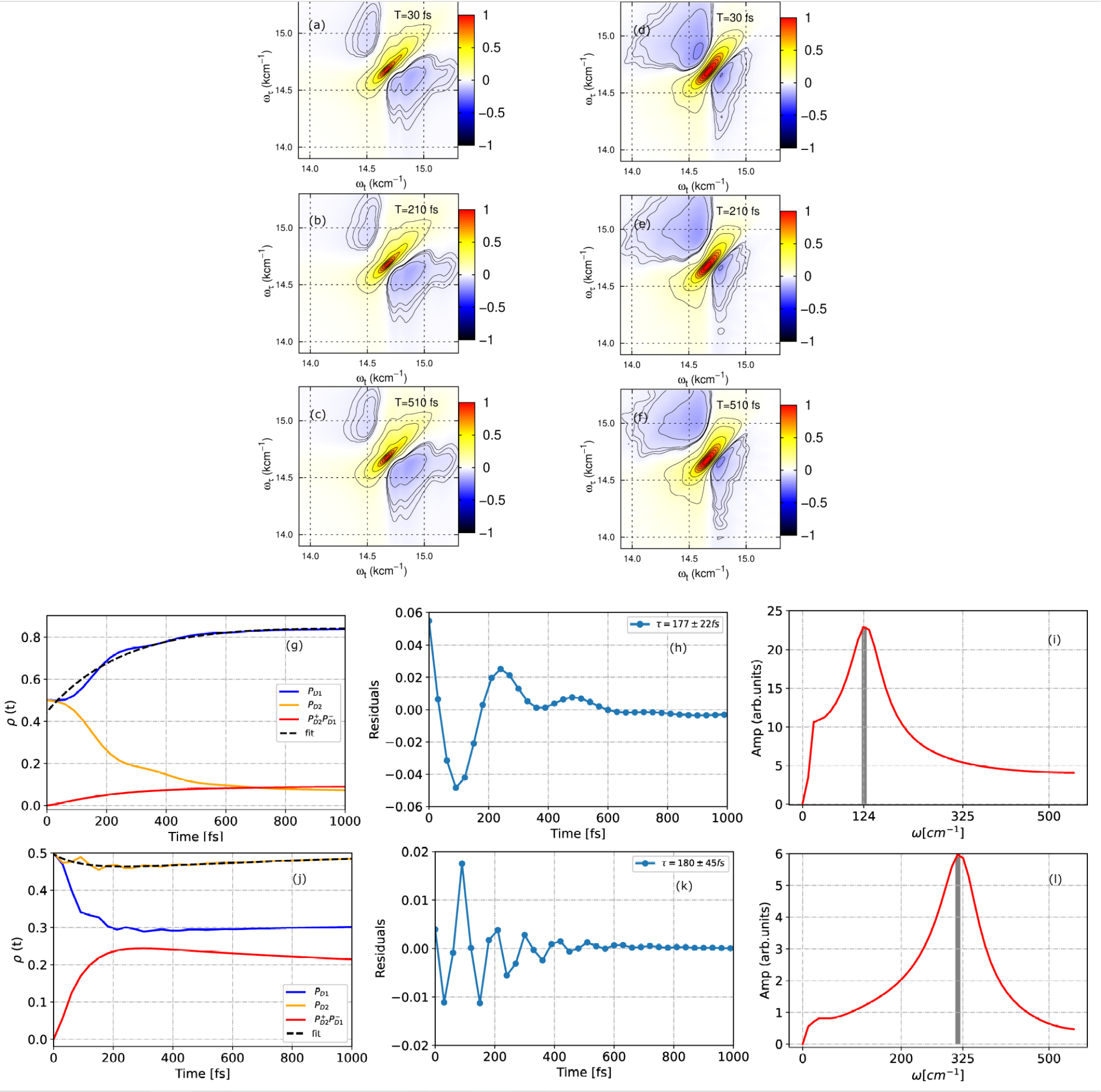} 
\caption{\label{fig:Fig5} The calculated 2D electronic spectra (real rephasing part) at waiting times of 30, 210 and 510 fs in (a), (b) and (c), respectively. The experimental counterparts are shown from (d) to (f). Part (g) depicts the calculated population dynamics of $\rm Chl_{D1}$ and $\rm Pheo_{D1}$ with the CT state $\rm Chl^{+}_{D1}Pheo^{-}_{D1}$. The residuals and their Fourier transform  are plotted in (h) and (i). The calculated population dynamics of the radical pair, $\rm P_{D1}$ and $\rm P_{D2}$ are shown in (j) with the CT state $\rm P^{+}_{D2}P^{-}_{D1}$. The fitted results and residuals are shown in (k). The Fourier transform of residuals are depicted in (l). } 
\end{center}
\end{figure}

We present the calculated 2D electronic spectra (rephasing real part) of the PSII reaction center at the temperature of 20 K in Fig.\ \ref{fig:Fig5}(a) to (c). The experimental counterparts are shown in Fig.\ \ref{fig:Fig5}(d) to (f), respectively. On the basis of our modeling and our refined fitting of the parameters, we observe that the calculated 2D spectra at different waiting times are in excellently agreement with the measured 2D spectra. Especially, the bandwidth of the anti-diagonal profile agrees well with measured data. This degree of agreement clearly illustrates that the parameters of the system-bath interaction in our model capture the decay time of the electronic dephasing between ground and excited states in the reaction center. The major discrepancy between theory and experiment arises from the ESA features in the upper-left side of the 2D electronic spectra. Clearly, more complicated ESA transitions are involved which are not included in the modeling and the calculations of the 2D spectra. It could be more precisely captured by the methods based on quantum chemistry calculations \cite{Thomas2011}. However, the current exciton model of the PSII reaction center captures the main features in the 2D electronic spectra. Thus, our model and the refined parameters are sufficient to compare to the experimental results. 

The modified Redfield theory used here is based on the second-order approximation of the system-bath interaction and cannot provide numerically exact results for the calculation of the electronic quantum coherence. Thus, we also employ the quasi-adiabatic propagator path integral (QUAPI) method \cite{Nancy1, Nancy2, Michael1} to examine the quantum coherent dynamics. We start from  our model and its optimized parameters obtained from 2D spectroscopic calculations. We then construct a dimer model for the study of the energy and CT dynamics. For this, we build the first dimer model of the pigments $\rm Chl_{D1}$ and $\rm Pheo_{D1}$. In addition, a CT state $\rm Chl^{+}_{D1}Pheo^{-}_{D1}$ is included in the modeling. The detailed Hamiltonian and parameters are shown in the SI. We show the calculated results in Fig.\ \ref{fig:Fig5}(g).  The population dynamics of the $\rm Chl_{D1}$ and $\rm Pheo_{D1}$ states are plotted as red and blue solid lines. The  primary CT is presented as black dashed line in Fig.\ \ref{fig:Fig5}(g). 

Moreover, we performed an the exponential fit to get the residuals. The fitting curve is shown as magenta dots. The subsequent residuals are plotted as red solid line in Fig.\ \ref{fig:Fig5}(h). The lifetime of the oscillations is obtained by fitting to the exponentially decaying harmonic functions.  The results yield the decay time of the coherence of 177 fs$\pm$22 fs, which perfectly agrees with  the lifetime revealed in the experiment (180 fs). The detailed fitting and calculation procedures are shown in the SI.  To obtain the oscillation frequency, we further performed a Fourier transform of the residuals and plot the result in Fig.\ \ref{fig:Fig5}(i). It yields the frequency of 124 cm$^{-1}$, which perfectly agrees with the energy gap found in our 2D spectroscopic data (experimental results, magenta line in Fig.\ \ref{fig:Fig4}(a)). This calculation shows that the electronic coherence between $\rm Chl_{D1}$ and $\rm Pheo_{D1}$ has the frequency of 124 cm$^{-1}$ and the decay time of 177 fs at 20 K. More interestingly, we do not observe any evidence of electronic coherence in the CT state $\rm Chl^{+}_{D1}Pheo^{-}_{D1}$, which is plotted as the black dashed line in Fig.\ \ref{fig:Fig5}(g). The absence of electronic coherence in the state $\rm Chl^{+}_{D1}Pheo^{-}_{D1}$ is due to the large intrinsic dipole moment of the CT state, which yields to a strong system-bath interaction (strong reorganization energy, $>$1000 cm$^{-1}$). Thus, we conclude that the observed long-lived oscillatory dynamics in the 2D spectra (340 cm$^{-1}$, red solid line in Fig.\ \ref{fig:Fig4}(b)) purely originates from vibrational coherence (which is commonly long-lived). 

Next, we study the coherent dynamics of the radical pair $\rm P_{D1}$ and $\rm P_{D2}$. For this, we construct a dimer model with a CT state $\rm P^{+}_{D2}P^{-}_{D1}$. The model and parameters are presented in the SI. We show the calculated population dynamics in Fig.\ \ref{fig:Fig5}(j).  The population dynamics of $\rm P_{D1}$ and $\rm P_{D2}$ are plotted as red and blue solid lines.  In addition, the population of the CT state is shown as a black dashed line in Fig.\ \ref{fig:Fig5}(j). We also apply the fitting procedure to obtain the residuals, the quality of fit is illustrated by magenta dots. The residuals are plotted as red solid line in Fig.\ \ref{fig:Fig5}(k) and the results after Fourier transformation are shown in Fig.\ \ref{fig:Fig5}(l), respectively.  We then perform the fitting procedure to resolve the lifetime of the electronic coherence for which we obtain 180$\pm$45 fs.  The detailed fitting and the analysis are shown in the SI. By this, we resolve the electronic coherence between radical pair ($\rm P_{D1}$ and $\rm P_{D2}$) with  325 cm$^{-1}$ and the decay time of 180 fs at 20 K.  Moreover, we do not observe any evidence of electronic coherence in the CT state $\rm P^{+}_{D2}P^{-}_{D1}$. Our calculated results agree well with our experimental observations for the radical pairs in the PSII reaction center.  In particular, we conclude that the observed long-lived coherence observed in the measured data originate from vibrations shown in Fig.\ \ref{fig:Fig4}(c) (the red solid line). 

We further examine the robustness of the electronic quantum coherence against temperature. For this, we calculate the population dynamics of the dimer model for varying temperature while the other parameters are kept constant. The results are shown in Fig.\ S9 in the SI for the dynamics of $\rm Chl_{D1}$ and $\rm Pheo_{D1}$, which is calculated at 80 K. It is interesting that we do not observe any evidence for oscillatory dynamics in the population transfer between $\rm Chl_{D1}$ and $\rm Pheo_{D1}$. In addition, the population dynamics of the CT state $\rm Chl^{+}_{D1}Pheo^{-}_{D1}$ is plotted as black dashed line in Fig.\ S9. It neither shows  evidence of electronic coherence. Moreover, we further calculate the population dynamics at room temperature (300 K), with the results being shown in Fig.\ S11 in the SI. We observe no signature of coherent dynamics in the calculated results. 

However, the results for the racial pair, $\rm P_{D1}$ and $\rm P_{D2}$, show a completely different picture. We repeat the same procedure for the calculations and vary the temperature of 80 K and 300 K. The resulting data are plotted in Figs.\ S10 and S12 in the SI, respectively. At 80 K, interestingly, the coherent dynamics between $\rm P_{D1}$ and $\rm P_{D2}$ lasts for 250 fs. However, the coherent oscillations are absent in $\rm P^{+}_{D2}P^{-}_{D1}$ due to the strong system-bath interaction of the CT state. We further increase the temperature to 300 K and calculate the population dynamics. The resulting data are plotted in Fig.\ S12 in the SI. We observe that the coherent oscillations of the population dynamics in $\rm P_{D1}$ and $\rm P_{D2}$ are still present - even at room temperature. The electronic quantum coherence clearly persists for 120 fs, however, the oscillatory amplitudes are dramatically reduced at room temperature. In addition, the calculated results for the CT state ($\rm P^{+}_{D2}P^{-}_{D1}$) do not reveal any coherence during the population transfer. 

\section*{Discussions} 

On the basis of the theoretical calculations, we uncover the lifetime of the coherent dynamics in the dimer systems of $\rm Chl_{D1}$, $\rm Pheo_{D1}$ and $\rm P_{D1}$, $\rm P_{D2}$. By varying temperature, we observe that the electronic quantum coherence between  $\rm Chl_{D1}$ and $\rm Pheo_{D1}$ is fragile with respect to thermal fluctuations. The coherent dynamics is completely suppressed at 80 K. In contrast, coherent dynamics between $\rm P_{D1}$ and $\rm P_{D2}$ is found to be rather robust. It shows a weak oscillatory dynamics and lasts for 120 fs even at room temperature. The difference of both configurations is the excitonic coupling. In our model, there is a coupling of 46 cm$^{-1}$ between  $\rm Chl_{D1}$ and $\rm Pheo_{D1}$, which is in a range of intermediate coupling strength in a protein complex. In contrast, the excitonic coupling of $\rm P_{D1}$ and $\rm P_{D2}$ is 150 cm$^{-1}$, which is definitely in the range of strong coupling. Such a rather large value of an excitonic coupling is unique for a  reaction center protein complex. It is due to the short distance of 3 \AA between the two pigments. This short distance results in a strong overlap of electronic orbitals between the two molecules. In fact, this could be an effective approach to overcome the dissipation-induced dephasing of quantum coherence at room temperature. 

This conclusion may be further supported by the transition between $\rm Chl_{D1}$ and $\rm Pheo_{D1}$. The value of the excitonic coupling of 46 cm$^{-1}$ is a typical value for energy transfer in photosynthetic protein complex. With this relatively weak coupling strength, the electronic coherence is absent even at 80 K. However, the short distance between $\rm P_{D1}$ and $\rm P_{D2}$ results in the CT process, which is not necessary similar for the other types (energy transfer) of  photosynthetic protein complexes. In fact, sometimes, CT is detrimental for them. More importantly, we also unravel that, due to the strong system-bath interaction, the CT state lacks substantial electronic quantum coherence. Most likely, this is the reason for  the negligible role of electronic coherence in the CT and charge separations in photovoltaics. 

We further extend our model to study the role of vibrational coherence in the CT. For this, we supplement our dimer model by adding an underdamped mode in the Ohmic spectral density. We kept all other remaining parameters unchanged. The detailed formula and parameters of spectral density are presented in the Materials and Methods section and the SI. We perform the calculation by QUAPI and plot the time-resolved population dynamics in the SI. We depict the dynamics of $\rm Chl_{D1}$ and $\rm Pheo_{D1}$ in Fig.\ S13 in the SI. The results show a long-lived electronic coherence between $\rm Chl_{D1}$ and $\rm Pheo_{D1}$ at 20 K. However, there is no evidence of vibrational coherence during the population transfer. This, in principle, could be induced by the weak vibronic coupling and also by the off-resonance of the vibrational frequency (340 cm$^{-1}$) and the excitonic energy gap (122 cm$^{-1}$). Moreover, vibrational coherence is not present in the CT state. In Fig.\ S14, we show the population dynamics of $\rm P_{D1}$ and $\rm P_{D2}$ at 20 K. The long-lived oscillatory dynamics shows a mixture of electronic and vibrational coherences in the time-dependent population of  $\rm P_{D1}$ and $\rm P_{D2}$. However, due to the strong system-bath interaction, the oscillation amplitudes in the population of $\rm P^{+}_{D2}P^{-}_{D1}$ is quite weak. Moreover, the calculations are repeated by QUAPI at 80 K and 300 K. We show the calculated results in Fig.\ S15 to S18 in the SI. We observe that, in the case of $\rm Chl_{D1}$ and $\rm Pheo_{D1}$, the electronic coherence is still fragile with respect to temperature.  No solid evidence of electronic and vibrational coherence during the population transfer in $\rm Chl_{D1}$ and $\rm Pheo_{D1}$ is found. In contrast, we do observe  clear evidence of a mixing of electronic and vibrational coherences in the population dynamics of  $\rm P_{D1}$ and $\rm P_{D2}$ at 80 and 300 K. In addition, vibrational coherence is also present in the CT state $\rm P^{+}_{D2}P^{-}_{D1}$, but with a quite weak magnitude. 

On the basis of these calculations, we conclude that, although the system-bath interaction is sizable, the vibrational coherence could survive in the CT states, yet with a relatively small magnitude. This perfectly agrees with our experimental observations in Fig.\ \ref{fig:Fig4}(b) and (c). Furthermore, our 2D spectroscopic measurements and theoretical calculations unravel that the decay time of the electronic quantum coherence is 177 fs during the primary CT ($\rm Chl_{D1}$ and $\rm Pheo_{D1}$) at 20 K. In addition, our theoretical calculations uncover the lifetime of electronic quantum coherence between $\rm P_{D1}$ and $\rm P_{D2}$ to be 180 fs. Due to the more blue spectral region of the exciton states of $\rm P_{D1}$ and $\rm P_{D2}$, our measurements cannot effectively reveal the electronic coherence between them due to the red-shift of our excitation laser spectrum (details are given in Fig.\ \ref{fig:Fig1}(b)). Moreover, we also calculate the population dynamics of the PSII reaction center, with the results shown in Fig.\ S19 and S20 in the SI. The results for  80 and 300 K show  clear evidence of electronic coherence of the states $\rm P_{D1}$ and $\rm P_{D2}$. Due to the much smaller thermal motions and resulting incresed memory time at 20 K, the applicability of the numerically exact QUAPI tool is limited by computational resource in the case of the population dynamics of the PSII reaction center. 

\section*{Conclusions} 

In this paper, we studied the coherent dynamics of the energy and charge transfer in the PSII reaction center complex by 2D electronic spectroscopy. The spectroscopic measurements at low temperature, namely 20 K, allow us to discern evidence of quantum coherence and to disentangle the electronic coherence from long-lived vibrational coherence. In the primary CT process, our measurements uncover a quantum coherent process with a timescale of 177 fs between $\rm Chl_{D1}$ to $\rm Pheo_{D1}$. Moreover, the modeling and careful refinement of the parameters used to calculate the 2D electronic spectra, enabled the capture of the main features of the experimental data. This analysis demonstrates the validity of the model and optimized parameters. We further employ the numerically exact QUAPI method to calculate the quantum coherent dynamics of the energy and CT processes. The results reveal the electronic coherence between $\rm Chl_{D1}$ to $\rm Pheo_{D1}$, and,  more importantly, the so obtained lifetime agrees well with the experimental measurements (177 fs). Interestingly, we also uncover the electronic coherence between $\rm P_{D1}$ and $\rm P_{D2}$, with the lifetime of 180 fs at 20 K. The theoretical calculations and spectroscopic measurements unravel that the vibrational coherence could persist for a rather long time during the CT dynamics despite the strong dissipation the CT states are exposed to due to the protein environment. We also examine the robustness of electronic and vibrational coherences at different elevated  temperatures. We uncover that, due to the closed configuration of pigments, the strong excitonic interaction between $\rm P_{D1}$  and $\rm P_{D2}$ results in a strong electronic coherence which even could survive at room temperature. However, the electronic coherence induced by an excitonic interaction in the region of intermediate coupling is fragile in view of the rather strong dissipation of the environment. Based on these observations, we conclude the correct picture is one involving dissipation-driven CT in the PSII reaction center, which prevails over a coherence-driven transfer dynamics. The pathways of the population transfer are delineated by strong couplings between pigments and the protein environment and the downhill CT directed by dissipative coupling to the bath as encoded in the significant redshift of site energies of the CT states. Due to these common features, we believe that this conclusion can be extended to other, more complicated photosynthetic protein complexes as a general principle. 

%


\section*{Materials and Methods}
\subsection{Sample preparation.} 

Thylakoid membranes were isolated from A. thaliana plants as described in Ref.\ \citeonline{EMBO_J_28_3052_(2009)} till the centrifugation step at 6000 g. Thylakoid membranes were solubilized with 0.6\% dodecyl-D-maltoside (DDM) at a final chlorophyll concentration of 0.5mg/ml. The sucrose density ultracentrifugation was used to obtain PSII core particles as described in Ref. \citeonline{Biochem_43_9467_(2004)}. The purification of PSII RCs from PSII core particles proceeds as follows: the PSII core particles were diluted in BTS200 buffer (20 mM Bis Tris pH 6.5, 20 mM $\rm MgCl_2$, 5 mM $\rm CaCl_2$, 10 mM $\rm MgSO_{4}$, 0.03\% DDM, 0.2M sucrose) to a chlorophyll concentration of 0.15mg/ml and solubilized with an equal volume of 10\% Triton X-100 in BTS200 buffer for 20 min; then the material was loaded on a HiTrap Q Sepharose HP 1ml column (GE Healthcare) and washed with a BTS buffer until the eluate became colorless. Finally, the PSII RC particles were eluted from the column with 75 mM $\rm MgSO_4$ in a BTS200 buffer. 

\subsection{2D Electronic measurements with experimental conditions.} 

Details of the experimental setup have been described in earlier reports from our group \cite{Duan2017}. Briefly, the measurements have been performed on a diffractive optics based on an all-reflective 2D spectrometer with a phase stability of $\lambda/160$. The laser beam from a home-built nonlinear optical parametric amplifier (NOPA, pumped by a commercial femtosecond Pharos laser from Light Conversion) is compressed to $\sim$20 fs using the combination of a deformable mirror (OKO Technologies, 19 channels) and a prism pair (F2 material). Frequency-resolved optical grating (FROG) measurement is used to characterize the temporal profile of the compressed beam and the obtained FROG traces are evaluated using a commercial program FROG3 (Femtosecond Technologies). A obtained broadband spectrum carried a linewidth of $\sim$100 nm (FWHM) cantered at 680 nm, which covered the electronic transitions to the first excited state. Three pulses are focused on the sample with the spot size of $\sim$200 $\mu$m and the photon echo signal is generated at the phase-matching direction. The photon-echo signals are collected using Sciencetech spectrometer model 9055F which is coupled to a CCD linear array camera (Entwicklungsb{\"u}ro Stresing). The 2D spectra for each waiting time T were collected by scanning the delay time $ \tau = t_{1}-t_{2}$ in the range of [-450 fs, 250 fs] with a delay step of 2 fs. At each delay step, 100 spectra were averaged to reduce the noise ratio. The waiting time $ T = t_{3}-t_{2}$ was linearly scanned in the range of 2.0 ps with steps of 15 fs. For the probing of the secondary charge separation, the detection window has been extended to 400 ps. For all measurements, the energy of the excitation pulse is attenuated to 3 nJ with 1 kHz repetition rates. Phasing of the  obtained 2D spectra was performed using an ``invariant theorem", which has been described in Ref.\ \citeonline{JCP_115_6606_2001}. 

\subsection{Theoretical calculations.} 

A Frenkel-exciton model is constructed to calculate the coherent energy and charge transfer dynamics and the 2D electronic spectra of the PSII reaction center complex. The total Hamiltonian is constructed in the form of  system, bath and system-bath interaction terms, $H = H_{S}+H_{B}+H_{SB}$. We use a tight-binding model for the charge transfer in the reaction center. In terms of the creation (annihilation) operators $\hat{e}^{\dagger}_{m}$ ($\hat{e}_{m}$) and $\hat{h}^{\dagger}_{m}$ ($\hat{h}_{m}$) for an electron in the LUMO or a hole in the HOMO of the pigment $m$, the Hamiltonian reads 
\begin{eqnarray}
\begin{aligned}\label{eq:RC_TB}
 H_{S} &= \sum_{m,n}t^{e}_{mn}\hat{e}^{\dagger}_{m}\hat{e}_{n} + \sum_{m,n}t^{h}_{mn}\hat{h}^{\dagger}_{m}\hat{h}_{n} + \sum^{m\neq n}_{m,n}W^{d}_{mn}\hat{e}^{\dagger}_{m}\hat{h}^{\dagger}_{m}\hat{h}_{n}\hat{e}_{n} \\
 -& \sum_{m,n}V^{eh}_{mn}\hat{e}^{\dagger}_{m}\hat{h}^{\dagger}_{n}\hat{h}_{n}\hat{e}_{m} 
 + \frac{1}{2}\sum^{m\neq n}_{m,n}V^{e}_{mn}\hat{e}^{\dagger}_{m}\hat{e}^{\dagger}_{n}\hat{e}_{n}\hat{e}_{m} \\
 +& \frac{1}{2}\sum^{m\neq n}_{m,n}V^{h}_{mn}\hat{h}^{\dagger}_{m}\hat{h}^{\dagger}_{n}\hat{h}_{n}\hat{h}_{m} \\
 +& \frac{1}{4}\sum^{k\neq m}_{k,m}\sum^{l\neq n}_{l,n}K_{kl,mn}\hat{e}^{\dagger}_{k}\hat{h}^{\dagger}_{l}\hat{e}^{\dagger}_{m}\hat{h}^{\dagger}_{n}\hat{h}_{n}\hat{e}_{m}\hat{h}_{l}\hat{e}_{k}\, .
\end{aligned}
\end{eqnarray}
Here, $t^{e}_{mn}$ and $t^{h}_{mn}$ are the hopping matrix elements connecting  the LUMO and HOMO of different pigments, respectively. $W^{d}_{mn}$ describes the strength of the Coulomb interaction between the charges on the pigments $m$ and $n$ and  $V^{eh}_{mn}$ is the electron-hole interaction between the charges on the pigments  $m$ and $n$. Likewise, $V^{e}_{mn}$ and $V^{h}_{mn}$ are the electron-electron and hole-hole Coulomb repulsion, respectively. The last term $K_{kl,mn}$ represents the interaction energy of the permanent dipole moments of the two doubly excited states. The bath Hamiltonian can be written as $H_{B} = \sum_{j}\frac{\omega_{j}}{2}(\hat{p}^{2}_{j}+\hat{x}^{2}_{j})$, where $\hat{p}_{j}$ and $\hat{x}_{j}$ are the momentum and position of the $j$th harmonic bath mode. Moreover, they are linearly coupled to the electron and hole orbitals by the system-bath term, $H_{SB} = -\sum_{m}\sum_{j}[ \omega_{j}d^{e}_{mj}\hat{x}_{j}\hat{e}^{\dagger}_{m}\hat{e}_{m} + \omega_{j}d^{h}_{mj}\hat{x}_{j}\hat{h}^{\dagger}_{m}\hat{h}_{m} ]$. The coupling strengths of system and bath are determined by the parameters of $\hat{F}^{e}_{m}=-\sum_{j}\omega_{j}d^{e}_{mj}\hat{x}_{j}$ and $\hat{F}^{h}_{m}=-\sum_{j}\omega_{j}d^{h}_{mj}\hat{x}_{j}$. The distribution of the vibrational frequencies of the bath is specified by an Ohmic spectral density $J_m(\omega) = \eta\omega\exp(-\omega/\omega_{c})$. To study the vibrational coherence, we include an underdamped mode in the spectral density which generates the total spectral density $J(\omega) = \eta\omega\exp(-\omega/\omega_{c}) + \frac{4S\gamma_{\mathrm{vib}}\omega_{\mathrm{vib}}^3\omega}{\left(\omega^2-\omega_{\mathrm{vib}}^2\right)^2+4\gamma_{\mathrm{vib}}^2\omega^2}$. Here, $\eta$ and $\omega_{c}$ are the damping strength and the cutoff frequency of the Ohmic spectral density, respectively. $S$, $\omega_{\mathrm{vib}}$ and $\gamma_{\mathrm{vib}}^{-1}$ are the Huang-Rhys factor, the vibrational frequency and the vibrational relaxation time of the underdamped mode, respectively. 

The nonequilibrium dynamics of the system-bath model is calculated by a modified Redfield quantum master equation, the details of which are described in the SI. The linear response theory is used to calculate the absorption spectrum of the reaction center complex 
\begin{eqnarray}
\begin{aligned}\label{eq:RC_abs}
 I(\omega)=\left\langle\int_0^{\infty}dte^{i\omega{t}}\mathrm{tr}(\bm{\mu}(t)\bm{\mu}(0)\rho_g)\right\rangle_{\mathrm{rot}} , 
\end{aligned}
\end{eqnarray}
where $\rho_g=|g\rangle\langle{g}|$ and a $\delta$-shaped laser pulse is assumed.  $\langle\cdot\rangle_{\mathrm{rot}}$ denotes the rotational average of the molecules with respect to the laser direction. Moreover, the 2D electronic spectra are obtained by calculating the third-order response function 
\begin{equation}
S^{(3)}(t,T,\tau)=\left(\frac{i}{\hbar}\right)^3\Theta(t)\Theta(T)\Theta(\tau)\mathrm{tr}\left(\bm{\mu}(t+T+\tau)\left[\bm{\mu}(T+\tau),\left[\bm{\mu}(\tau),\left[\bm{\mathrm{\mu}}(0),\rho_g\right]\right]\right]\right). 
\end{equation}
Here, $\tau$ is the delay time between the second and the first pulse, $T$ (the so-called waiting time) is the delay time between the third and the second pulse, and $t$ is the detection time. To evaluate 2D electronic spectra, we need the rephasing (RP) and non-rephasing (NR) contributions of the third-order response function, i.e., $S^{(3)}(t,T,\tau)=S_{\mathrm{RP}}^{(3)}(t,T,\tau)+S_{\mathrm{NR}}^{(3)}(t,T,\tau)$. Assuming the impulsive limit (the $\delta$-shaped laser pulse), one obtains 
\begin{eqnarray}
I_{\mathrm{RP}}(\omega_t,T,\omega_{\tau})&=&\int_{-\infty}^{\infty}d\tau\int_{-\infty}^{\infty}dt e^{i\omega_{t}t-i\omega_{\tau}\tau}S_{\mathrm{RP}}^{(3)}(t,T,\tau), \\
I_{\mathrm{NR}}(\omega_{t},T,\omega_{\tau})&=&\int_{-\infty}^{\infty}d\tau\int_{-\infty}^{\infty}dt e^{i\omega_{t}t+i\omega_{\tau}\tau}S_{\mathrm{NR}}^{(3)}(t,T,\tau).
\end{eqnarray} 
The total 2D signal is the sum of the two, i.e., $I(\omega_{t},T,\omega_{\tau})=I_{\mathrm{RP}}(\omega_{t},T,\omega_{\tau})+I_{\mathrm{NR}}(\omega_t,T,\omega_{\tau})$. 

The model parameters of the site energies and electronic couplings are directly taken from Ref. \citeonline{ChemPhysChem}. The strengths of the electronic couplings are taken without further change but the site energies are refined during a simultaneous fit to the absorption of the PSII reaction center complex at different temperatures. The final versions of the site energies are very close to the version reported in Ref. \citeonline{Jennifer2013, Jennifer2014}. To precisely determine the reorganization energy, the parameters are further refined by fitting to the experimental anti-diagonal bandwidth of the main peak in 2D spectra.

%
\begin{addendum}
\item We thank Roberta Croce and Henny van Roon for providing PSII reaction center sample. We thank V. I. Prokhorenko for help with the 2D setup and for providing the 2D data analysis software. This work was supported by NSFC grant with NO.\ 12274247 and the foundation of national excellent young scientist. The Next Generation Chemistry theme at the Rosalind Franklin Institute is supported by the EPSRC (V011359/1 (P)) (AJ). This work was also supported by the Max Planck Society and by the Cluster of Excellence ‘Advanced Imaging of Matter’, EXC 2056, Project ID 390715994 of the Deutsche Forschungsgemeinschaft. 

\item[Supporting information] The supplementary Information includes the analysis of the lifetime of optical dephasing obtained from the anti-diagonal bandwidth, the description of the global fitting approach and the obtained DAS, and the Tukey window Fourier transform and the disentangling of electronic coherence from vibrations by Curve Fitting Toolbox. Moreover, it describes the 2D power spectra of the identified vibrational modes, the basis transformation between site and excitonic states, the model and parameters of PSII reaction center and the model of the dimer ($\rm P_{D1}$, $\rm P_{D2}$ and $\rm Chl_{D1}$, $\rm Pheo_{D1}$). Finally, we review the quantum master equation of the modified Redfield approach and the numerically exact method of QUAPI. 

\item[Competing Interests] The authors declare that they have no competing financial interests. 

\item[Correspondence] Correspondence of paper should be addressed to H.-G.D. ~(email: duanhongguang@nbu.edu.cn), M.T. ~(email:michael.thorwart@physik.uni-hamburg.de) and R.J.D.M.~(email: dmiller@lphys.chem.utoronto.ca) 

\item[Author contributions] H. -G. D. conceived the research and discussed with A. J., M. T. and R. J. D. M.. H.-G. D., A. J. and V. T. performed the spectroscopic measurements. P.-P. Z., H.-G. D., and L. C. constructed the model and performed theoretical calculations. H.-G. D., A. J., M. T. and R. J. D. M. wrote the initial draft and refined by all authors. H.-G. D., M. T. and R. J. D. M. supervised this project. 

\end{addendum}
%
\newpage

\newpage

\newpage

\newpage

\newpage


\begin{thebibliography}{11}
 
\bibitem {RC_crystal} Umena, Y., Kawakami, K., Shen, J. R. \& Kamiya, N. Crystal structure of oxygen-evolving Photosystem II at a resolution of 1.9\AA. Nature \textbf{473}, 55–60 (2011). 

\bibitem {RC_crystal2} Dods, R. {\em et al.} Ultrafast structural changes within a photosynthetic reaction center. Nature \textbf{589}, 310-314 (2020). 

\bibitem {RC_crystal3} Suga, M. {\em et al.} Native structure of photosystem II at 1.95 $\rm \AA$ resolution viewed by femtosecond X-ray pulses. Nature \textbf{517}, 99 (2015). 

\bibitem {D1_branch} Diner, B. A. \& Rappaport, F. Structure, dynamics, and energetics of the primary photochemistry of Photosystem II of oxygenic photosynthesis. Annu. Rev. Plant Biol. \textbf{53}, 551–580 (2002). 

\bibitem {Novoderezhkin_I} Novoderezhkin, V. I. {\em et al.} Pathways and Timescales of Primary Charge Separation in the
Photosystem II Reaction Center as Revealed by a Simultaneous Fit of Time-Resolved Fluorescence and Transient Absorption. Biophys. J. \textbf{89}, 1464-1481 (2005). 

\bibitem {Novoderezhkin_II} Novoderezhkin, V. I. {\em et al.} Mixing of Exciton and Charge-transfer states in Photosystem II Reaction centers: Modeling of Stark Spectra with Modified Redfield Theory. Biophys. J. \textbf{93}, 1293-1311 (2007). 

\bibitem {Valentyn} Prokhorenko, V. I. \& Holzwarth, A. R. Primary processes and structure of the Photosystem II reaction center: a photon echo study. J. Phys. Chem. B \textbf{104}, 11563–11578 (2000).

\bibitem {Biochemistry} Romero, E. {\em et al.} Two different charge separation pathways in Photosystem II. Biochem. \textbf{49}, 4300–4307 (2010). 

\bibitem {ChemPhysChem} Novoderezhkin, V. I. {\em et al.} Multiple Charge-separation pathways in photosystem II: Modeling of transient absorption kinetics. Chem. Phys. Chem. \textbf{12}, 681-688 (2011). 

\bibitem {Engel2007} Engel, G. S. {\em et al.} Evidence for wavelike energy transfer through quantum coherence in photosynthetic systems. Nature \textbf{446}, 782-786 (2007).

\bibitem {Scholes2010} Collini, E. {\em et al.} Coherently wired light-harvesting in photosynthetic marine algae at ambient temperature. Nature \textbf{463}, 644-647 (2010). 

\bibitem {Duan_Sci_Adv2020} Cao, JS, {\em et al.} Quantum biology revisited. Science Adv. \textbf{6}, eaaz4888 (2020). 

\bibitem {Duan2017} Duan, H. G. {\em et al.} Nature does not rely on long-lived electronic quantum coherence for photosynthetic energy transfer. Proc. Natl. Acad. Sci. (USA) \textbf{114}, 8493-8498 (2017). 

\bibitem {Scholes2017} Scholes, G. D. {\em et al.} Using coherence to enhance function in chemical and biophysical systems. Nature \textbf{543}, 647-656 (2017). 

\bibitem {Zigmantas2018} Thyrhaug. E. {\em et al.} Identification and characterization of diverse coherences in the Fenna–Matthews–Olson complex. Nature Chem. \textbf{10}, 780-786 (2018). 

\bibitem {Duan2022} Duan, H. -G. {\em et al.} Quantum coherent energy transport in the Fenna–Matthews–Olson complex at low temperature. Proc. Natl. Acad. Sci. (USA) \textbf{119}, e2212630119 (2022). 

\bibitem {Nature_rev_chem2019} Wang, L., Allodi, M. A. \& Engel, G. S. Quantum coherences reveal excited-state dynamics in biophysical systems. Nature Rev. Chem. \textbf{3}, 477-490 (2019). 

\bibitem {Romero2017} Romero, E., Novoderezhkin, V. I. \& Grondelle, van R. Quantum design of photosynthesis for bio-inspired solar-energy conversion. Nature \textbf{543}, 355–365 (2017). 

\bibitem {David2013} Tiwari, V., Peters, K. W., \& Jonas, D. M. Electronic resonance with anticorrelated pigment vibrations drives photosynthetic energy transfer outside the adiabatic framework. Proc. Natl. Acad. Sci. (USA) \textbf{110}, 1203-1208 (2013). 

\bibitem {Plenio2013} Chin, A. W. {\em et al.} The role of non-equilibrium vibrational structures in electronic coherence and recoherence in pigment–protein complexes. Nature Phys. \textbf{9},113-118 (2013). 

\bibitem {Miller2014} Halpin, A. {\em et al.} Two-dimensional spectroscopy of a molecular dimer unveils the effects of vibronic coupling on exciton coherences. Nature Chem. \textbf{6},196-201 (2014). 

\bibitem {Duan2015} Duan, H. -G. {\em et al.} On the origin of oscillations in two-dimensional spectra of excitonically-coupled molecular systems. New J. Phys. \textbf{17}, 072002 (2015). 

\bibitem {JPCL2012} Kreisbeck, C. \& Kramer, T. Long-Lived Electronic Coherence in Dissipative Exciton Dynamics of Light-Harvesting Complexes. J. Phys. Chem. Lett. \textbf{3}, 2828-2833 (2012). 

\bibitem {Dassia2014}  Egorova, D. Self-analysis of coherent oscillations in time-resolved optical signals. J. Phys. Chem. A \textbf{118}, 10259-10267 (2014). 

\bibitem {Hauer2013} Milota, F. {\em et al.} Vibronic and Vibrational Coherences in Two-Dimensional Electronic Spectra of Supramolecular J-Aggregates. J. Phys. Chem. A \textbf{117}, 6007-6014 (2013).  

\bibitem {Jennifer2010} Myers, J. A. {\em et al.} Two-dimensional electronic spectroscopy of the D1-D2-cyt b559 photosystem II reaction centre complex. J. Phys. Chem. Lett. \textbf{1}, 2774-2780 (2010).

\bibitem {Jennifer2012} Lewis, K. L. M. {\em et al.} Simulation of the two-dimensional electronic spectroscopy of the photosystem II reaction center. J. Phys. Chem. A \textbf{117}, 34-41 (2013). 

\bibitem {Jennifer2013} Gelzinis, A. {\em et al.} Tight-binding model of the photosystem II reaction center: application to two-dimensional electronic spectroscopy. New J. Phys. \textbf{15}, 075013 (2013). 

\bibitem {Romero2014} Romero, E. {\em et al.} Quantum coherence in photosynthesis for efficient solar-energy conversion. Nature Phys. \textbf{10}, 676-682 (2014). 

\bibitem {Jennifer2014} Fuller, F. D. {\em et al.} Vibronic coherence in oxygenic photosynthesis. Nature Chem. \textbf{6}, 706-711 (2014).  

\bibitem {Fleming2022} Yoneda, Y. {\em et al.} The initial charge separation step in oxygenic photosynthesis. Nature Comm. \textbf{13}, 2275 (2022). 

\bibitem {Jennifer2021} Song, Y. {\em et al.} Excitonic structure and charge separation in the heliobacterial reaction center probed by multispectral multidimensional spectroscopy. Nature Comm. \textbf{12}, 2801 (2021). 

\bibitem {global fitting} Prokhorenko, V. I. Global analysis of multi-dimensional experimental data.Eur. Photochem. Assoc. Newslett. June21 (2012). 

\bibitem {DuanJPCB2015} Duan, H. G. {\em et al.} Two-dimensional electronic spectroscopy of light-harvesting complex II at ambient temperature: A joint experimental and theoretical study. J. Phys. Chem. B \textbf{119}, 12017-12027 (2015). 

\bibitem {DuanSciRep_2017} Duan, H. G. {\em et al.} Primary charge separation in the photosystem II reaction center revealed by a global analysis of the two-dimensional electronic spectra. Scientific Rep. \textbf{7}, 12347 (2017). 

\bibitem {PNAS1995} Peterman, E. J. {\em et al.} The nature of the excited state of the reaction center of photosystem II of green plants: A high-resolution fluorescence spectroscopy study. Proc. Natl. Acad. Sci. (USA) \textbf{95}, 6128-6133 (1998). 

\bibitem{Duan_PRE_2015} Duan, H. G., Dijkstra, G. A., Nalbach, P. and Thorwart, M. Efficient tool to calculate two-dimensional optical spectra for photoactive molecular complexes. Phys. Rev. E \textbf{92}, 042708 (2015). 

\bibitem{Fleming2002} Yang, M. and Fleming, G. R. Influence of phonons on exciton transfer dynamics: comparison of the Redfield, Forster, and modified Redfield equations. Chem. Phys. \textbf{282}, 163 (2002). 

\bibitem{JCP1998} Zhang, W. M., Meier, T., Chernyak, V. Exciton-migration and three-pulse femtosecond optical spectroscopies
of photosynthetic antenna complexes. J. Chem. Phys. \textbf{108}, 7763 (1998). 

\bibitem{Thomas2011} Olbrich, C., {\em et al.} From atomistic modeling to excitation transfer and two-dimensional spectra of the FMO light-harvesting complex. J. Phys. Chem. B \textbf{115}, 8609 (2011). 

\bibitem {Nancy1} Makri, N. \& Makarov, D. E. Tensor propagator for iterative quantum time evolution of reduced density matrices. I. Theory. J. Chem. Phys. \textbf{102}, 4600 (1995). 

\bibitem {Nancy2} Makri, N. \& Makarov, D. E. Tensor propagator for iterative quantum time evolution of reduced density matrices. II. Numerical methodology. J. Chem. Phys. \textbf{102}, 4611 (1995). 

\bibitem {Michael1} Nalbach, P., Braun, D., \& Thorwart, M. Exciton transfer dynamics and quantumness of energy transfer in the Fenna-Matthews-Olson complex. Phys. Rev. E \textbf{84}, 041926 (2011). 

\bibitem {EMBO_J_28_3052_(2009)} Caffarri, S. {\em et al.} Functional architecture of higher plant photosystem II supercomplex. EMBO. J. \textbf{28}, 3052-3063 (2009). 

\bibitem {Biochem_43_9467_(2004)} Caffarri, S. {\em et al.} A look within LHCII: differential analysis of the Lhch1-3 complexes building the major trimeric antenna complex of higher-plant photosynthesis. Biochemistry \textbf{43}, 9467-9476 (2004). 

\bibitem {JCP_115_6606_2001} Hybl, J. D. Ferro, A. A. \& Jonas, D. M. Two-dimensional Fourier transform electronic spectroscopy. J. Chem. Phys. \textbf{115}, 6606 (2001). 
 
\end{thebibliography}
\end{document}